\documentclass[twocolumn,prl,showpacs,floatfix]{revtex4}
\usepackage{graphicx}
\usepackage{dcolumn}
\usepackage{amsmath}
\usepackage{latexsym}
\begin{document}

\title{Dynamical Transitions of a Driven Ising Interface}

\author{Manish K. Sahai and Surajit Sengupta}
\affiliation{
Satyendra Nath Bose National Centre for Basic Sciences, Block-JD, 
Sector-III, Salt Lake, Kolkata 700 098, India
}
\date{\today}
\begin{abstract}
We study the structure of an interface in a three dimensional Ising system 
created by an external non-uniform field $H({\bf r},t)$. $H$ changes 
sign over a two dimensional plane of arbitrary orientation. When the field 
is pulled with velocity ${\bf v}_e$, 
(i.e. $H({\bf r},t) = H({\bf r - v_e}t)$), the interface undergoes a 
several dynamical transitions. For low velocities it is pinned by the 
field profile and moves along with it, the distribution of local slopes 
undergoing a series of commensurate-incommensurate transitions. For large 
${\bf v}_e$ the interface de-pinns and grows with KPZ exponents. 
\end{abstract} 
\maketitle

{\em Introduction:\,} The study of the structure and properties of surfaces 
and interfaces in condensed matter is of considerable technological 
importance\cite{intro1,intro2}. Many phase transitions and chemical 
reactions originate at a surface and their progress 
in time is often synonymous with the motion of resulting parent- product 
interfaces\cite{intro3}. Often, interfaces undergo structural transitions, like 
reconstructions and roughening, the study of which have recieved considerable 
attention in the past because of its importance to a variety of processes like
catalysis and surface phase transitions\cite{intro1,villain}. Surface 
structure can also be influenced by its dynamics. Consider a simple interface 
between the ground states of an Ising model\cite{chaikin} with nearest 
neighbor ferromagnetic interactions defined on the $d$ dimensional cubic 
lattice. The equilibrium, static, interface defined by the height $h({\bf r})$ 
fluctuates so that the mean squared deviation of $h$ from the average 
grows\cite{barabasi} as $L^{\alpha}$ and $t^{\beta}$ where $L$ is the typical
system size, $t$ the time with the exponents $\alpha = (2-d)/2$  and
$\beta = 2 -d/4$. Above $d=2$ the interface remains flat 
($\alpha = \beta = 0$). On the other hand, a {\em moving} interface in this 
system when driven by an uniform field coarsens such that 
$\alpha \simeq 2/(d+3)$ and $\beta \simeq 1/(d+2)$. 

\begin{figure}[h]
\begin{center}
\includegraphics[width=8.0cm]{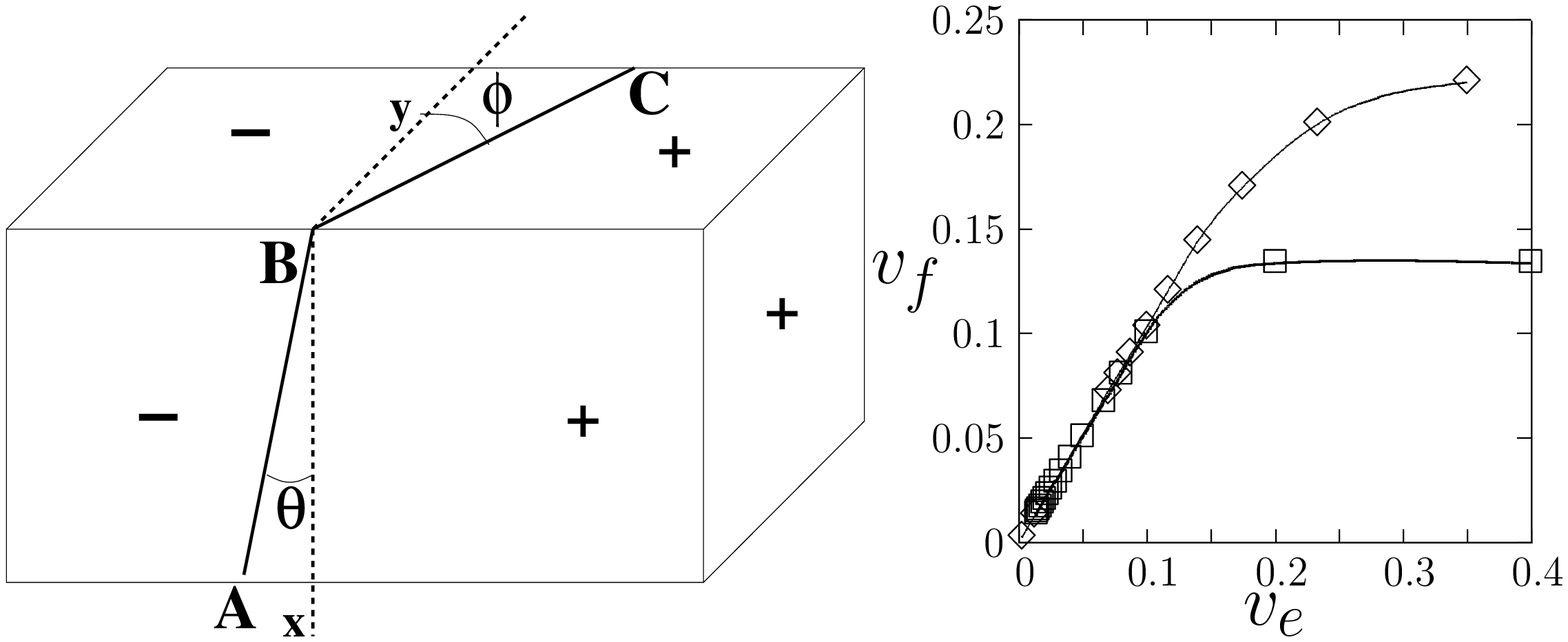}
\end{center}
\caption{(left) The geometry of our simulation cell. Ising spins are defined
over a $14 \times 14 \times 30$ lattice elongated in the $z$ direction. An
external field is imposed which changes sign over a plane $A,B,C$ making
angles $\theta$ and $\phi$ with the $x$ and $y$ axes respectively. The plane
$A,B,C$ moves to the right with velocity $v_e$. (right) A plot of the velocity
of the Ising interface $v_f$ as a function of $v_e$ for two different
orientations -  (0,2/7)(top) and (2/7,2/7) . Note that for small $v_e$, the 
interface follows the field
($v_f = v_e$) but gets detached at higher velocities.
}
\label{one}
\end{figure}

{\em Motivation:\,} In a few earlier publications\cite{physica,prl,pt} we have 
shown that it is, nevertheless, possible to stabilize a 
{\em macroscopically flat} ($\alpha = \beta = 0$) but moving interface in this 
system by using a {\em non-uniform} field (see Fig.\ref{one}) of the form 
$H({\bf r},t) = 
H_0 {\rm tanh}[({\bf r} - {\bf v}_e t)\cdot \hat{\bf n}/\chi]$ where the 
unit vector $\hat{\bf n}$ fixes the orientation of the interface
and $\chi$ the (intrinsic) width. In Ref.\cite{prl,pt} the properties of 
such an interface in the {\em two dimensional} Ising model was analyzed in 
detail.        
The main results of this study were as follows. Firstly, there exists two 
distinct dynamical phases: for small $v_e$, the interface, defined as the
locus of all the points where the magnetization $S({\bf r},t)$ changes sign,  
moves together with the field with velocity $v_f = v_e$, and 
its average position is fixed at all times by the plane ({\em line} in two 
dimensions) $z_0(x,y)$ over which the field changes sign and the interface 
remains macroscopically flat. For large $v_e > v_{\infty}$, on the other 
hand, the interface detaches from $z_0$ and grows 
with velocity $v_{\infty}$ essentially driven by an uniform field at the 
same time coarsening with KPZ exponents\cite{KPZ,barabasi}. The value of 
$v_{\infty}$ is 
orientation dependent because the motion of interfaces at low 
temperatures takes place mainly by flipping the relatively high energy spins 
occupying corner sites. The interfacial orientation with the largest density 
of such unstable spins has the highest $v_{\infty}$.   

Secondly, the low velocity 
pinned phase was shown to be rather interesting because it shows an 
{\em infinite} number of structural phases and phase transitions very similar
to commensurate-incommensurate transitions in adsorbates on solid 
surfaces\cite{chaikin,FK,CI}. 
In the pinned phase, the spatial profile of the field forces the magnetization 
to follow $H$ as closely as possible. Specifically, the average orientation of 
the interface is fixed by the orientation of the line $z_0$ over which 
$H$ changes sign.  Since the underlying lattice admits only a discrete set of
rational orientations, at equilibrium, the interface settles down to the
closest rational approximant possible given the finite size of the system.
The pinning energy depends on the orientation of the interface relative to
the lattice, simple rational fractions being more strongly pinned. 
Increasing $v_e$ has the effect of increasing dynamical noise, and orientations
with stronger pinning and shorter relaxation times become preffered. As
$v_e$ is increased, therefore, an interface with an arbitrary orientation 
distorts locally so that over most of its length, the orientation conforms to 
a low order rational fraction. A non-zero density  of 
{\em discommensurations}\cite{chaikin,prl,pt} maintains the average orientation of the 
interface equal to that of
$z_0$. A plot of the most probable orientation of the interface against
the average orientation for any $v_e < v_{\infty}$ therefore shows steps
corresponding to an incomplete Devil's staircase 
structure\cite{chaikin,FK,prl,pt}.

The purpose of the present Brief Report, is to determine whether these 
conclusions are valid for the $d=2$ interface in the $d=3$ Ising system and are 
not an artifact of the reduced dimensionality. Most experimentally relevant 
interfaces being two dimensional, we believe this to be an essential point
which needs to be checked by direct calculation. 
We show that the general scenario persists even in the
higher dimensional case studied here, although there are some special features
related to the larger degree of freedom of the two dimensional interface
compared to the one dimensional case studied in Refs.\cite{prl,pt}. Unlike
the former study, where one may make use of a formal mapping of the Ising
interface dynamics problem to an Assymetric Exclusion Process\cite{barma} to
develop efficient numerical codes and obtain useful analytical results, no
such mappng is possible for the higher dimensional case and we use
standard Monte Carlo simulations using Metropolis update for this problem.
The exact nature of the update rule is not expected to influence the
qualitative nature of our results.
\begin{widetext}
.
\begin{figure}[h]
\begin{center} \includegraphics[width=15.0cm]{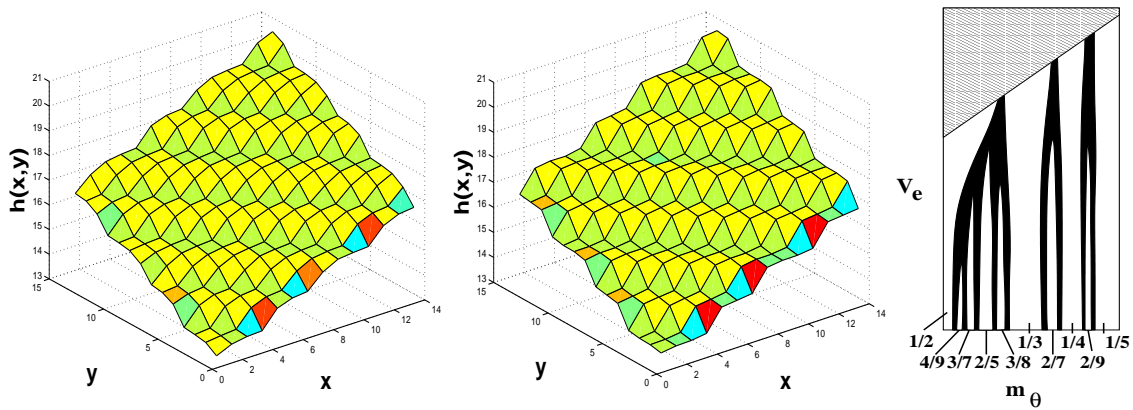}
\end{center} 
\caption{(left (a) and middle (b)) Plots of the averaged interface defined by 
the 
locus of zeroes of the magnetization for two different values of 
$v_e = .1$ (left) and $.0125$. The orientation of the interface is ($2/7,2/7$).
Note that for small smaller $v_e$ the interface is more sharply defined.
(right (c) ) A portion of the dynamical phase diagram for a {\em one dimensional}
interface taken from Ref.\cite{pt} showing locked (white regions), fluctuating
(black regions) and de-pinned (hatched region) phases. Note that an interface
with orientation close to (and to the right of) $2/7$ would pass through 
regions where the most probable orientation shifts to ($1/4$) as the $v_e$ is 
increased.
}
\label{intepha}
\end{figure}
\end{widetext}

{\em Mean field results:\,} When spatial flucutations of the interface are 
neglected, the velocity of the interface $v_f$, defined as the locus of all the 
points where the magnetization $S({\bf r},t)$ changes sign, is given by the 
solution of $v_f = v_{\infty} {\rm sign}(v_f - v_e)$ at late times\cite{prl}. 
There are 
two regimes: (a) for small $v_e < v_{\infty}$ the interface is pinned by the 
field and its position coincides with the plane over which the field $\phi$ 
changes sign so that $v_f = v_e$ and (b) for $v_e > v_{\infty}$, the interface
depins and grows with its orientation dependent intrinsic velocity 
$v_{\infty}$. These mean field results are independent of dimensionality and 
should hold for Ising systems in any dimensions. We show, indeed, that this is 
true and the depinning transition, predicted by the mean field calculation 
persists in $3-d$ (see Fig.\ref{one}).  

{\em Simulation details:\,}Our system consists of 
$N_x (= 14) \times N_y (= 14) \times N_z (= 30)$ Ising spins arranged in a 
simple cubic lattice\cite{lanbin} interacting ferromagnetically with their nearest 
neighbors. The boundary conditions for this system has to be specified with 
some care since we need to describe, in general, oblique interfaces with 
arbitrary orientations. We are interested in the low temperature, high field 
limit, however, which greatly simplifies this task. In this limit, only a few 
layers of spins immediately adjacent to the interface have any nontrivial 
dynamics, spins far away from the interface being frozen in the direction of 
the local external field $H$. We use {\em shifted} periodic 
boundary conditions ensuring that the periodic image of each spin has the same 
value of $H$ (see Fig 1.). In the ${\bf z}$ direction we have open boundary 
conditions the exact nature of which is irrelevant as the spins at the two 
extremeties are, in any case, frozen to the values fixed by the sign of $H$.

A typical run consists of equilibrating the spins in a given $H$, updating 
$H$ according to the value of $v_e$ measured in units of lattice parameter
per Monte Carlo step (MCS) and computing the non-uniform magnetization. The 
value of the magnetization is stored for each time step till the interface 
traverses the length of the sample. This process is typically repeated $10^3$ 
times and the averaged time dependent magnetization profile is obtained.
To obtain the location of the interface the averaged magnetization
$S({\bf r},t)$ is fitted to a general hyperbolic tangent profile. The locus of 
zeroes is the interface. The probability distribution of slopes of this 
interface is then computed using finite differences and is further averaged 
over $10^2$ independent runs. Therefore, although our system size of 
$5880$ spins is modest, the scale of our computations is not; a run 
with a single parameter value needs to be repeated many times to gather 
sufficient statistics.  
 
{\em Results and Discussion:\,}We begin our discussion of results by 
considering first the velocity of the interface $v_f$ as a function of $v_e$. 
We obtain this from the slope of a plot of the position of the center of mass 
of the interface as a function of time. Our results for two different 
orientations of $\hat{\bf n}$ are shown in Fig. 1 (right panel). The depinning
transition and saturation of $v_f$ to a orientation dependent value of 
$v_{\infty}$ is clearly visible. The saturation value $v_{\infty}$ depends
on the density of steps which, in turn, is a function of the interfacial 
orientation. Compared to the two dimensional case of Refs.\cite{prl,pt}, 
however, this 
transition appears to be less sharp. This is, in fact, a finite size
effect\cite{prl,lanbin}; the sharp transition is recovered only in the large 
time limit which 
is difficult to achieve for our modest system size. Note that the transition 
is broader for the faster moving interface for which the observation time is 
shorter.   

We concentrate next on the interfacial structure of the pinned interface. The 
average orientation of the interface is given by the unit vector 
$\hat{\bf n}$ which may be parametrized by the angles $\theta$ and $\phi$ as 
shown in Fig. 1 (left panel). We find it more convenient to define the slopes 
$m_{\theta} = \tan(\theta)$ and $m_{\phi} = \tan(\phi)$. The structure of the 
interface averaged over many runs is shown in Fig.\ref{intepha} (a) and (b) 
for two different velocities and for $m_{\theta} = m_{\phi} = 2/7$. 
It is clear from these pictures that the velocity strongly influences the 
structure, smoothening it on the average. In order to understand better this 
smoothening process, we compute the probability distribution of the local 
and instantaneous slope using a resolution window of $5 \times 5$ spins. 
Within this window, the local slope is determined at every intance of time and 
the data is used to build up the joint probability distribution 
$P(m_{\theta},m_{\phi})$.    

\begin{figure}[t]
\begin{center} \includegraphics[width=8.5cm]{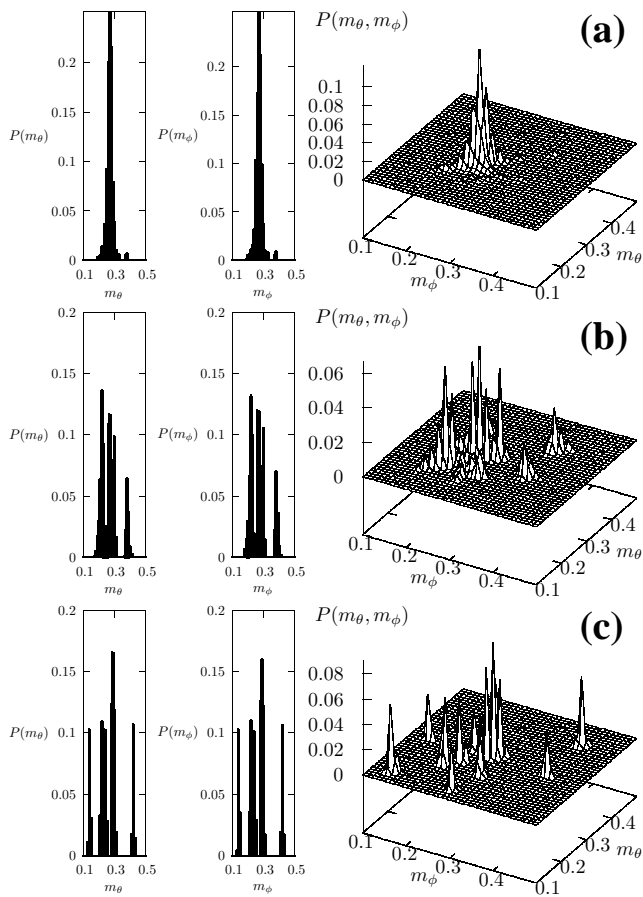}
\end{center} 
\caption{
Projected and two dimensional distributions of the local slope of the interface
with an average slope close to $m_{\theta} = 2/7$ and $m_{\phi} = 2/7$ for 
various values of $v_e$. Left panel shows the probability distribution of 
the slope $P(m_{\theta})$, the middle panel $P(m_{\phi})$ and the right panel
the full two dimensional distribution $P(m_{\theta},m_{\phi})$ for 
$v_e = .2$ (a), $.1$ (b), and $.0125$ (c). Note that in (a), the most probable 
slope is close to the average slope of the interface. As the velocity is 
lowered, the interface tends to lock in to a rational slope close to $1/4,1/4$, 
for which the probability shows the strongest peak (b). As the
$v_e$ is lowered further, the lock-in slope is a higher order fraction 
$2/7,2/7$ which is closer to the average orientation of the interface. 
Large fluctuations away from the lock-in slope are, however, present 
throughout contributing secondary peaks in the probability distribution 
function.  
}
\label{config}
\end{figure}

The joint probability distribution $P(m_{\theta},m_{\phi})$ as well as the 
projected distributions $P(m_{\theta})$ and $P(m_{\phi})$ are plotted in 
Fig.\ref{config} for three different $v_e$. In order to understand the 
sequence of the transitions, we refer to a portion of the dynamical phase 
diagram in the $v_e - m_{\theta}$ plane Fig.\ref{intepha}(c). 
In this figure, the white regions represents the lock-in phases where the 
most probable orientation locks in to the closest rational slope which may be 
different 
from the average slope of the interface. The black regions represent 
incomensurate states where the the most probable slope is nearly equal to the 
average slope and the interface dynamically fluctuates between nearby 
orientations. Finally, the gray region for large $v_e$ corresponds to the 
depinned interface. An interface with a slope $m_{\theta}$ close to but 
smaller than $2/7$ would first lock-in at $2/7$ for small $v_e$ and then to 
$1/4$ and finally depinn as $v_e$ increases. In-between these transitions the 
interface would be incommensurate.    

This sequence of transitions is apparent in the plots of 
$P(m_{\theta},m_{\phi})$ shown in Fig.\ref{config}. However, in contrast
to the two dimensional situation study\cite{prl}, the interface 
in the three dmensional Ising system has more degrees of freedom. This is 
evidenced by the presence of multiple local orientations which show up as 
secondary peaks for small $v_e$. Careful observation of the structure of the 
interface shows that the local slope of the interface continually fluctuates
in space and time over a small region of phase space around the average 
orientation. 

What determines the strength of the pinning in any particular 
orientation? The residence time in any particular orientation depends both 
on the strength of the pinning as well as on the nearness in configuration 
space from the average orientation. In the study of the one dimensional 
interface, the pinning energies could be computed analytically by mapping 
the one dimensional interface to a sequence of particles in a one dimensional 
lattice\cite{barma,AEP}. This procedure is not possible in the present case. 
However, the conclusions 
of this study seems to be valid for each of the two slopes $m_{\theta}$ and 
$m_{\phi}$ taken separately. A simple ansatz which proceeds to break up the 
response of the $2-d$ interface with slope $(m_{\theta},m_{\phi})$ as that of 
two $1-d$ interfaces with slopes $m_{\theta}$ and $m_{\phi}$ succeeds fairly 
well. The lowest energy orientations correspond to $m_{\theta}$ or $m_{\phi}$ 
being of the form $1/n$ where $n$ is an integer. So the interface with 
an initial slope near to  $2/7,2/7$ (Fig.\ref{config}(c)) for the smallest 
velocity, breaks up into regions with slopes close to $2/7$ with the largest 
probability corresponding to a slope close to $1/4,1/4$ (Fig. \ref{config}(b)) 
which is the similar to that for the $1-d$ interface\cite{prl}. However, as 
mentioned before and as is clear from Fig. \ref{config}(b). 
there are considerable weights for nearby slopes where the interface spends 
a considerable amount of time. These appear as secondary peaks in 
$P(m_{\theta},m_{\phi})$. If the velocity is increased further, the interface 
becomes incommensurate and the probability distribution nearly Gaussian with 
a sharp peak only for the average slope (Fig. \ref{config}(a)). Further we 
have verified with actual computation that the 2-d probability distribution
is {\em not} given by a product of the 1-d probabilities making fluctuations 
in $m_{\theta}$ and $m_{\phi}$ highly correlated. 

A detailed analytical calculation of the full probability distribution would 
be useful but is beyond the scope of this work. Future work in this direction 
would hopefully explain these issues analytically and obtain the full three 
dimensional phase diagram in the $m_{\theta}, m{\phi}$ and $v_e$ plane. We
hope our studies will be useful in understanding moving interfaces in more 
realistic systems\cite{acmrss} and the growth of colloidal crystals using 
patterned substrates\cite{AMOLF}.  
 
\acknowledgements The authors thank A. Chaudhuri, A. Milchev and 
D. Dimitrov for discussions.  



\end{document}